\documentclass[runningheads]{llncs}

\usepackage{eccv}

\usepackage{eccvabbrv}

\usepackage{graphicx}
\usepackage{booktabs}
\usepackage{times}
\usepackage{epsfig}
\usepackage{graphicx}
\usepackage{bm}
\usepackage{bbm}
\usepackage{amssymb}
\usepackage{mathtools}
\usepackage{comment}
\usepackage{xcolor}
\usepackage{booktabs,makecell,multirow}
\usepackage{threeparttable}
\usepackage{colortbl}
\usepackage{makecell}
\usepackage{wasysym}
\usepackage{wrapfig}
\usepackage[ruled,linesnumbered]{algorithm2e}
\usepackage[T1]{fontenc}
\definecolor{dark-gray}{gray}{0.10}
\definecolor{light-gray}{gray}{0.9}
\definecolor{blue-light}{RGB}{245, 255, 255}
\definecolor{green-light}{RGB}{255, 255, 245}
\definecolor{gray-light-fig1}{gray}{0.95}

\usepackage[accsupp]{axessibility}  % Improves PDF readability for those with disabilities.

\usepackage{hyperref}

\begin{document}
\title{Revisiting Adaptive Cellular Recognition Under \\Domain Shifts: A Contextual Correspondence View\vspace{-3.8mm}} 

\titlerunning{Revisiting Adaptive Cellular Recognition Under Domain Shifts}

\author{
Jianan Fan\inst{1} \and
Dongnan Liu\inst{1} \and
Canran Li\inst{1} \and
Hang Chang\inst{2} \and
Heng Huang\inst{3} \and\\
Filip Braet\inst{1} \and
Mei Chen\inst{4} \and
Weidong Cai\inst{1}
\vspace{-0.5mm}
}

\authorrunning{F.~Author et al.}

\institute{University of Sydney, Camperdown, NSW 2050, Australia \and
Lawrence Berkeley National Laboratory, Berkeley, CA 94720, USA \and
University of Maryland at College Park, College Park, MD 20742, USA \and
Microsoft, Redmond, WA 98052, USA}
% Springer Heidelberg, Tiergartenstr.~17, 69121 Heidelberg, Germany

\maketitle

\begin{abstract}
\vspace{-4.2mm}
  Cellular nuclei recognition serves as a fundamental and essential step in the workflow of digital pathology. However, with disparate source organs and staining procedures among histology image clusters, the scanned tiles inherently conform to a non-uniform data distribution, which induces deteriorated promises for general cross-cohort usages. Despite the latest efforts leveraging domain adaptation to mitigate distributional discrepancy, those methods are subjected to modeling the morphological characteristics of each cell individually, disregarding the hierarchical latent structure and intrinsic contextual correspondences across the tumor micro-environment. In this work, we identify the importance of implicit correspondences across biological contexts for exploiting domain-invariant pathological composition and thereby propose to exploit the dependence over various biological structures for domain adaptive cellular recognition. We discover those high-level correspondences via unsupervised contextual modeling and use them as bridges to facilitate adaptation over diverse organs and stains. In addition, to further exploit the rich spatial contexts embedded amongst nuclear communities, we propose self-adaptive dynamic distillation to secure instance-aware trade-offs across different model constituents. The proposed method is extensively evaluated on a broad spectrum of cross-domain settings under miscellaneous data distribution shifts and outperforms the state-of-the-art methods by a substantial margin. Code is available at https://github.com/camwew/Cellular-Recognition\_DA\_CC.
  %\keywords{Cellular recognition \and Domain adaptation \and Contextual Correspondence}
\end{abstract}

\section{Introduction}
\begin{wrapfigure}{r}{0.55\columnwidth}
	\centering
	\vspace{-8mm}
	\includegraphics[width=0.55\columnwidth]{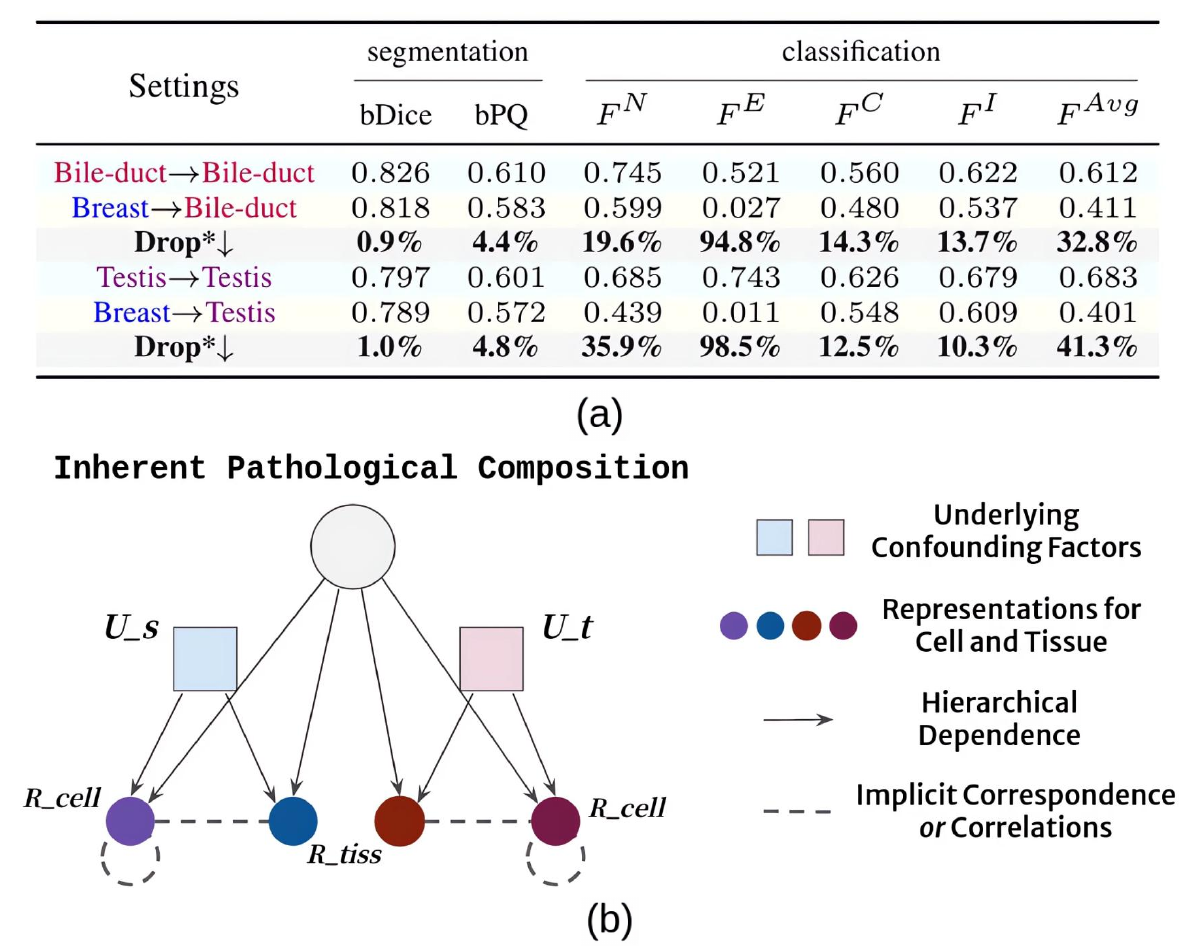}
	\caption{
		(a)\;Illustrative results for cellular recognition under domain shifts. 
		X$\rightarrow$Y denotes that the model is trained with data from X organ and then evaluated on Y organ. 
		bDice and bPQ measure the accuracy of class-agnostic segmentation. 
		$F^*$ denote the $F$ score for different nuclear types, indicating subtyping accuracy. 
		(b)\;Schematic diagram of the hierarchical nature of latent variables.
		The pathological composition principle of tumor micro-environment inherits fundamental \emph{invariance} regardless of the underlying confounding factors such as sampling organs and staining protocols, holding great promises to formalize domain-agnostic biomarkers.
	}
	\vspace{-2mm}
	\label{fig:tissue-wise}
\end{wrapfigure}

%\begin{figure}[!t]
%	\centerline{\includegraphics[width=1.1\columnwidth]{tissue-wise-performance-eccv.pdf}}
%	\caption{
%		(a)\;Illustrative results for nuclei recognition under domain shifts. 
%		X$\rightarrow$Y denotes that the model is trained with data from X organ and then evaluated on Y organ. 
%		bDice and bPQ are metrics to measure the accuracy of class-agnostic segmentation. 
%		$F^*$ denote the $F$ score for different nuclear types, indicating classification accuracy. 
%		(b)\;Schematic diagram of the hierarchical nature of latent variables and representations.
%		The pathological composition principle of tumor micro-environment inherits fundamental invariance regardless of the underlying confounding factors such as sampling organs and staining protocols, demonstrating great promises to formalize domain-agnostic biomarkers.
%	}
%	\label{fig:tissue-wise}
%	\vspace{-3mm}
%\end{figure}
\label{sec:introduction}
%Digital pathology, which aims to assist clinicians with automated quantitative analysis of biological structure expression and tissue sections, has recently shown great promise in clinical practice \cite{nagpal2019development, ghahremani2022deepliif}.
In the routine of pathological examination, cellular recognition, which aims to identify the specific type of each cell nucleus and segment it,
%(e.g., neoplastic, non-neoplastic epithelial, connective, and inflammatory) 
%plays an essential role in the characterization of tumor micro-environment \cite{lu2018nuclear, wu2022cross}. 
contributes fundamentally to the success of various downstream clinical commissions \cite{lu2018nuclear, wu2022cross}.
%, such as cancer prognosis 
%\cite{rodemann2011functional} 
%and survival prediction \cite{lu2018nuclear, saha2018her2net}.
Despite the numerous learning-based approaches proposed in this context \cite{tyagi2023degpr, huang2023affine, he2023toposeg}, those efforts are founded upon an ill-considered assumption that histopathology imaging data is uniformly distributed across cohorts, which dismiss the potential collapse incurred by cross-organ/stain variances \cite{gamper2019pannuke, graham2021lizard}.
In digital pathology practices, query samples could be acquired from divergent organs with inconsistent staining procedures \cite{kumar2019multi, verma2021monusac2020}.
The inherently present data distribution shifts induce far-reaching detriment to the robustness and general applicability of learned recognition model.
The quantitative illustration and evidence are showcased in Fig.\;\ref{fig:tissue-wise}(a).
It is observed that the precision of cellular subtyping degrades drastically when evaluated across domains, though the performance of class-agnostic segmentation is relatively robust.

A promising solution is to leverage well-annotated data from one source cohort and then perform unsupervised domain adaptation\,(UDA) \cite{liu2020unsupervised, hsu2021darcnn, fan2023taxonomy} to transfer the learned domain-agnostic knowledge to another data collection with disparate distribution.
However, the efficacy of existing UDA methods in the context of cross-domain cellular recognition is intrinsically undermined by the following limitations:
\textbf{i)}\;Most of these efforts are subjugated to aligning low-level visual attributes in a category-agnostic manner \cite{xing2020bidirectional, yang2021minimizing, fan2024learning}.
%Whilst in regard to nuclei recognition and classification, 
%dismissing the substantial texture pattern discrepancy between the same category of nuclei across domains.
Without exploitation of discriminative contextual characteristics along adaptation, there is no guarantee that cells of different subtypes can be well separated under domain shifts.
\textbf{ii)}\;In spite of the recent endeavors to perform class-conditioned alignment \cite{zheng2020cross, li2022domain, li2022sigma},
a major deterrent stands that they either depend on feature representations from the source domain or require pseudo-labeling to categorize objects in the target domain. 
%and then align instance-level features in a class-wise manner.
In digital pathology, the large discrepancy between the source and target domain at the feature distribution level and the unreliability of pseudo-labels on cellular species would inevitably lead to error accumulation and biased alignment.
\begin{wrapfigure}{r}{0.55\columnwidth}
	\centering
	\vspace{-8mm}
	\includegraphics[width=0.55\columnwidth]{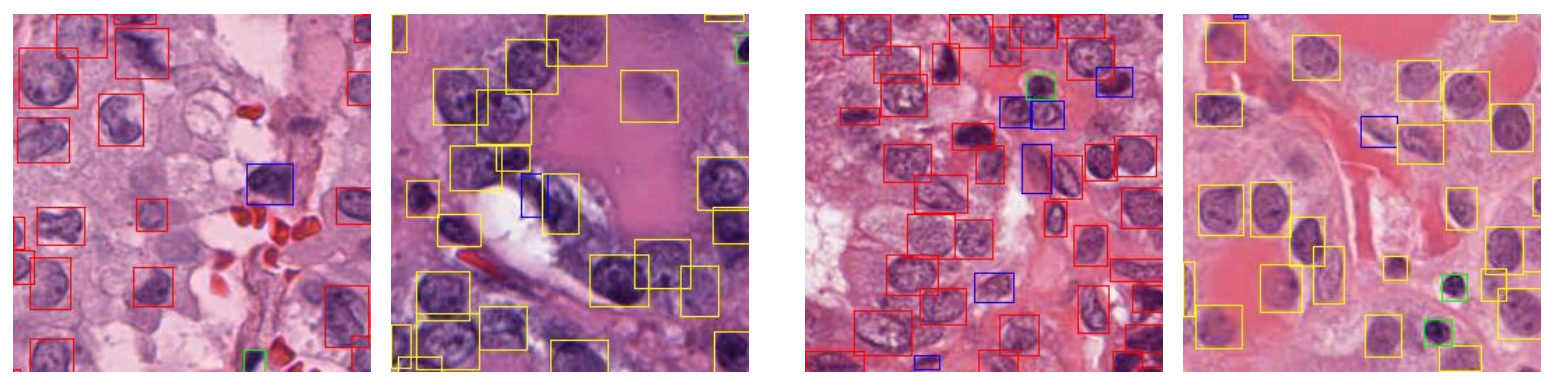}
	\caption{
		Exemplary H\&E-stained histology tiles. 	
		In each sub-figure, red, yellow, blue, and green rectangles correspond to the nuclei of neoplastic, epithelial, connective, and inflammatory cells, respectively.
		% The tissue structures of neoplastic cells hold intricate texture patterns compared with those of epithelial cells. 
		%and can be adopted as robust markers for ambiguous nuclei recognition.
	}
	\label{fig:inter-class}
	\vspace{-2mm}
\end{wrapfigure}
\textbf{iii)}\;Most importantly, those approaches seek to devise instance-level alignment strategy to transfer object semantics, which indicate the morphological characteristics\;(\emph{e.g.}, appearance, texture, and shape) of each individual object, to facilitate adaptation across domains \cite{chen2018domain, zhou2022multi, cao2023contrastive}.
However, cellular recognition differs from general perception tasks in computer vision by large contexts.
Specifically, other than object semantics, the underpinning \emph{pathological composition principle and the resultant relational correspondences across biological structures}
%\footnotemark\footnotetext{Note that the word ``tissue'' in ``background tissue'' and ``cross-tissue'' have different meanings. The former one indicates the biological structures surrounding nuclei, while the latter one denotes the organ where the images are sampled from.}
are also indispensable for distinguishing visually ambiguous cells.
As illustrated in Fig.\;\ref{fig:tissue-wise}(b), we characterize the composing process of tumor micro-environment with a hierarchical formulation between latent variables and representations.
With conjoint parents in the directed acyclic graph \cite{peters2017elements}, the representations for nuclear clusters and adjacent tissues are causally subjected to statistical dependence.
The implicit correspondences among latent representations also instigate observable implications on biological structure traits.
As Fig.\;\ref{fig:inter-class} shows, the tissue structures surrounding neoplastic and epithelial cells demonstrate significant visual differences,
despite the analogous morphological properties of nuclei themselves. 
It suggests that jointly modeling cells and surrounding tissues and exploiting their correspondences are advantageous for recognizing cells with ambiguous semantics \cite{ryu2023ocelot}.
Furthermore, the biological correlations concurrently hold that the spatial contexts among neighboring nuclei in nuclear clusters are found to be informative and could behave as discriminative markers for cell identification \cite{abousamra2021multi}.
In this work,
we propose a novel framework harnessing the inherent biological correspondences across pathological environment for cross-domain nuclei recognition.
Our insight lies in leveraging the correspondences between observable biological structures to formulate a suite of surrogate tasks, from which the \emph{underlying pathological composition principle} can be implicitly learned and exploited.
The unveiled pathogenesis principle inherits coherence regardless of the confounding factors such as sampling organs and staining procedures, and can therefore behave as \emph{domain-invariant knowledge}.
By feeding neural models with the explicit invariance, we present an elegant solution to overcome shortcut learning biased from spurious correlations, which has been regarded as the primary determinant for model collapse under domain shifts \cite{saranrittichai2022overcoming}.
%By injecting
% and facilitate model generalization.
In specific,
we devise a multifaceted self-discovery scheme to 
%firstly 
uncover the correspondences across biological structures over complementary levels and 
%thereafter
intrinsically learn the coherent data-genesis principle to endow the adapted model with general serviceability.
It is achieved by devising 
%interventional 
pretext tasks to perform counterfactual nuclei masking for exploring the implicit 
% nuclei-tissue and inter-nuclei
dependencies in an unsupervised manner.
% and then distilling the learned implicit relational knowledge to endow the model with general transferability.
%The intuition is further illustrated in Fig.\;\ref{fig:crosstask}.
%First, we mask all the nuclei regions in histology image patches and then add an auxiliary branch to reconstruct the missing pixel values based on the remaining background tissue.
%Then, to discover inter-nuclei correlations, we mask the RoI\;(\emph{i.e.}, Region of Interest) feature maps of the nucleus at the center of each image patch and then propose to recover the instance features according to other neighbouring nuclei with a Transformer \cite{dosovitskiy2021image}-based module.
Moreover, we propose self-adaptive dynamic distillation to further leverage the correspondences within nuclear clusters via exploiting the spatial contexts of neighbouring nuclei and accordingly conduct instance-adaptive rectifications on model outputs.
%To this end, we add a classification head on top of the Transformer encoder and propose an uncertainty-aware cross-prediction consistency module to evaluate the trade-offs between the two predictions and thereby provide object-adaptive guidance for Mask R-CNN.
%We induce the model to generate two parallel results and propose an uncertainty-aware cross-prediction consistency module on them to estimate the trade-offs between predictions and provide object-adaptive guidance.
%These auxiliary tasks and modules have task-specific decoders, while they share the backbone network with the primary framework (i.e., Mask R-CNN) for cross-task knowledge distillation.
%Furthermore, they are employed simultaneously for both source and target tissues to ensure category-wise nuclei characteristics in both domains are modeled.

%\begin{figure}[!t]
%	\centerline{\includegraphics[width=0.9\columnwidth]{inter-class-eccv.pdf}}
%	\caption{Exemplary H\&E-stained histology tiles. 
%		%(at 40× objective magnification, $\sim$0.25\textmu m/pixel).
%		%of thyroid tissue.
%		% from the PanNuke dataset \cite{gamper2019pannuke}. 
%		In each sub-figure, red and yellow rectangles correspond to the nuclei of neoplastic and epithelial cells, respectively.
%		It can be observed that the tissue structures of neoplastic cells hold intricate texture patterns compared with those of epithelial cells and can be adopted as robust markers for ambiguous nuclei recognition.}
%	\label{fig:inter-class}
%	\vspace{-3mm}
%\end{figure}
In a nutshell, our contributions are three-fold:
\textbf{i)}\;We develop a novel framework for cross-domain cellular nuclei recognition which, for the first time, goes beyond ambiguous object semantics and proposes to leverage the latent hierarchy in the pathological composing process and the implicit biological correspondences as bridges to foster model adaptation.
\textbf{ii)}\;We propose the hierarchical self-discovery and instance-adaptive rectification methodologies to exploit the multifaceted biological correspondences for learning high-level composition principle via surrogate tasks, without the need for prior pathological knowledge.
\textbf{iii)}\;We comprehensively evaluate our proposed method and demonstrate its effectiveness on diverse cross-domain settings, attaining remarkable improvements over state-of-the-art UDA methods.
%It significantly outperforms state-of-the-art UDA methods for nuclei recognition.

\section{Related Work}

\noindent\textbf{UDA in Biomedical Images.} \ 
%The robustness of learning-based models is argued to be susceptive of the data distribution shifts across different domains \cite{zhou2022domain}.
Aiming at mitigating the distributional bias and learning transferable knowledge across data cohorts, unsupervised domain adaptation\;(UDA) \cite{zhou2022generalizable, shin2023sdc} is raised as a popular line of research.
Representative works for UDA propose to mitigate the discrepancy across domains from the image appearance and feature representation level.
To accomplish appearance-level alignment, a generative or style hallucinative model is introduced to transform source domain images to synthetic target-like ones \cite{murez2018image, liu2020unsupervised, zhang2022exact}. 
%Afterward, based on the generated target-like images and the annotations of source images, a target-specific model is trained to perform inference on target images \cite{xing2020bidirectional}.
%Furthermore, to enhance the unsupervised image translation model with additional supervision, several efforts have been made to design an auxiliary task-specific branch and integrate it with the primary image translation framework as regularization to encourage the modeling and transfer of task-related attributes \cite{zhang2018task, cai2019towards}. 
With respect to content representation-level alignment, distributional disparity regularization and adversarial optimization are commonly adopted approaches to derive domain-invariant representation \cite{chen2020unsupervised, wu2021unsupervised, zhang2022towards}.
Those methodologies have shown impressive efficacy for alleviating cross-domain gap in various scenarios such as anatomical structure segmentation \cite{han2021deep} and nucleus detection \cite{xing2020bidirectional}.

\noindent\textbf{Domain Adaptive Object Recognition.} \ 
In the context of cross-domain object recognition, the prevailing approach is employing region-of-interest\;(ROI) based domain discriminators to align instance-level features from different domains \cite{chen2018domain}.
Following a similar spirit, multi-branch architectures are further devised to compensate for the model collapse issue and resort to securing the best trade-offs between domain-specific and domain-invariant attributes \cite{deng2023harmonious, he2023multi}.
Style hallucination techniques are also utilized to mitigate domain discrepancy from the image appearance level \cite{huang2021rda}.
%the feature map outputs of a specific convolutional layer, referred to as global image-level feature alignment.
%Additionally, an domain classifier is also deployed for the RoI-based feature vectors, referred to as local instance-level feature alignment \cite{chen2018domain}.
However, the aforementioned approaches stand without consideration of category-related characteristics along domain alignment and therefore tend to incur negative transfer across irrelevant entity types.
In recent literature, several efforts \cite{zheng2020cross, zhou2022multi, kennerley20232pcnet} have posited the beneficial practice of performing class-wise alignment and harnessing category pseudo-labels from the target domain.
For instance, 
%\cite{zheng2020cross} models the semantic prototype of each category and minimizes the pairwise prototype distances across domains with pseudo-labels.
\cite{he2023multi} proposes a multi-scale adversarial training paradigm which jointly minimizes image-level domain discrepancy and aligns semantic representation across domains in a class-aware manner.
Nonetheless, considering the model vulnerability in cross-domain scenarios for the nuclei recognition task and the resulting biased pseudo-labels, securing precise class-wise alignment across distinct nuclei types is deemed impractical.
%\cite{vs2021mega} further proposes to construct feature memory bank for category-aware domain adaptation, yet it collects source domain features to update the memory bank and neglects the severe distribution shifts between feature representations from the source and target domains.
Analogous to our work, \cite{yang2021minimizing, li2022domain} study category-aware nuclei recognition under domain shifts. 
However, they adopt approaches similar to \cite{zheng2020cross} and are therefore suboptimal due to the lack of consideration for domain-invariant biological composition and the unstable pseudo-label learning process.

\begin{comment}
\subsection{Nuclei Recognition and Classification with Relationships}
Unlike the object recognition task in computer vision, where the object's semantics dominate the decision, the morphological attributes of nuclei alone are insufficient to derive accurate classification results.
Inspired by the clinical practice of pathologists, the spatial relationships with neighbouring nuclei and tissue architecture are identified as an essential factor for nuclei classification \cite{abousamra2021multi}.
Specifically, \cite{liu2021relational} employed a relational LSTM model to incorporate the contextual information of neighbouring cells in decision. 
\cite{abousamra2021multi} introduced the K-function \cite{bull2020combining} and designed a multi-task learning framework to learn a spatial-context-aware representation of cells.
\cite{hassan2022nucleus} adopted a graph-based algorithm to discover nuclear communities and refine nuclei classification.
However, those works only 
Despite the success those methods achieved in the fully-supervised setting, the potential of leveraging the inherent biological correlations is neglected under the cross-tissue scenario.
%Different from their approaches, we devise three auxiliary tasks and modules that can be integrated with the primary framework in an end-to-end manner and do not require additional annotations.
\end{comment}

\section{Methodology}
\subsection{A Hierarchical View on Pathology Data Genesis}
As a starting point, we characterize the underpinning pathological genesis process with a hierarchical latent variable model \cite{hinton2022represent}, as depicted in Fig.\;\ref{fig:tissue-wise}(b).
The hierarchical dependence structure can be formally described as follows:
\begin{proposition}[\textbf{Hierarchical Formulation of Latent Variables}]
	Let $\mathcal{G}^*$ be the directed acyclic graph describing the causal structure of latent variables, with the sets of nodes and edges denoted as $\mathcal{V}$ and $\mathcal{E}$.
	$\mathcal{V}$ is composed of the measurable representations $\mathbf{R}$ for biological structures as well as the high-level latent variables $\mathbf{S}$.
	Then, the underlying data genesis procedure can be characterized with the following structural
	formulations:
	$\mathbf{R}_i=\sum_{\mathbf{S}_j\in Pa(\mathbf{R}_i)}p_{ij}(\mathbf{S}_j)+Q(\epsilon_i)$,
	where $Pa(\cdot)$ represents the set of parents for a certain node, $p_{(\cdot,\cdot)}$ denotes the causal dependence function, and $Q(\epsilon)$ is the probability distribution over exogenous random variables $\{\epsilon_i\}$.
\end{proposition}
Based upon the principle, with the high-level pathological composition mechanism denoted as a latent variable $\mathbf{S}_{c}$, we have $\mathbf{S}_{c}$ as the conjoint ancestor for different biological structures, namely $\mathbf{R}_{c}\leftarrow\mathbf{S}_{c}\rightarrow\mathbf{R}_{t}$, where $\mathbf{R}_{c}$ and $\mathbf{R}_{t}$ stand for the representations of cells and tissues.
\begin{figure}[!t]
	\centerline{\includegraphics[width=0.98\columnwidth]{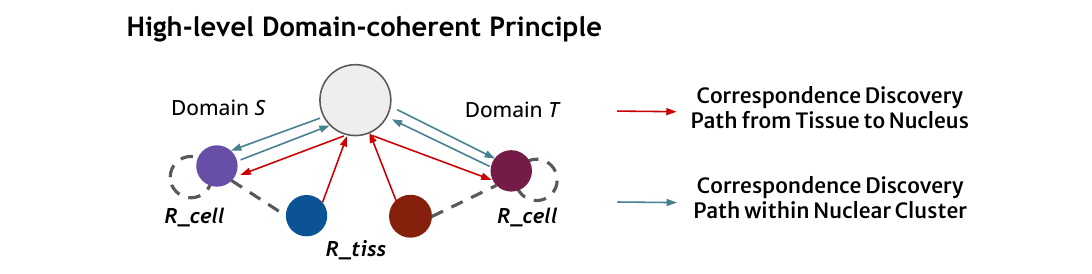}}
	\caption{Conceptual illustration of the insight. To exploit the intrinsic pathological composition principle which inherits cross-domain coherence, we propose to devise self-supervised surrogate tasks to discover multifaceted biological correspondences, from which the high-level principle variables can be implicitly learned to endow the model with strengthened generalizability.}
	\label{fig:idea}
	\vspace{-3mm}
\end{figure}
It implies that the \emph{correspondences between those biological structures can be traced back to the high-level latent factor} $\mathbf{S}_{c}$, which holds fundamental invariance across domains.
As Fig.\;\ref{fig:idea} illustrates, $\mathbf{S}_{c}$ serves as an intermediate node in the correspondence discovery chain\;(\emph{e.g.,} $\mathbf{R}_{t}\xmapsto[]{\mathbf{S}_{c}}\mathbf{R}_{c}$), such that the \emph{knowledge can be implicitly learned and injected to the model along the path}.
In this regard, we propose to implicitly learn the domain-coherent principle to foster model generalization by exploring the biological correspondences via counterfactual nuclei masking and restoration.

A schematic illustration of the proposed method is shown in Fig.\;\ref{fig:model}.
%Our intuition is to leverage self-supervised surrogate tasks endowed with cross-domain coherence to enable model transfer and refitting to the target domain, even without any tuning on the primary nuclei recognition task.
%By this means, we can get rid of the reliance on biased nuclei type pseudo-labels for model adaptation.
%The overall framework is composed of three critical constituents.
Specifically, we propose multifaceted correspondence self-discovery and instance-adaptive dynamic distillation, which aim to capture the inherent nuclei-tissue and nuclei-nuclei relationships and exploit relational contexts within nuclear clusters to rectify model predictions dominated by the characteristics of individual nucleus, respectively.
Those surrogate tasks share a backbone network with the primary nuclei recognition branch for seamless transfer of complementary knowledge.
%Following previous works \cite{hsu2021darcnn}, we adopt Mask R-CNN \cite{he2017mask} as the base model.
%The devised auxiliary tasks and modules are integrated with the Mask R-CNN framework for classification refinement.

\begin{comment}
\begin{figure}[!t]
\centerline{\includegraphics[width=0.9\columnwidth]{crosstask.png}}
\caption{Illustration of the insight to exploit domain-robust cross-task correlations for UDA. At first, an auxiliary task is specifically devised which is semantically related to the main task and consistent across domains. Then, the cross-task correlations are explored in the source domain, which is afterwards transferred to the target domain. In the end, we leverage the learned correlations as well as the auxiliary task trained in the target domain to serve as a bridge and thereby adapt the main task model.}
\label{fig:crosstask}
\end{figure}
\end{comment}
\subsection{Multifaceted Correspondence Discovery}
%\subsubsection{Correspondence across Nucleus and Tissue}
\noindent\textbf{Correspondence across Nucleus and Tissue.} \ 
As illustrated in Fig.\;\ref{fig:inter-class}, capturing the underlying correlations across nuclei and tissue structures offers discriminative contextual information and could benefit the precise identification of nuclear subtypes in spite of their ambiguous visual attributes.
Those intrinsic biological correspondences are rooted in the fundamental composition mechanism of pathology data and thus inherit stronger robustness against domain shifts compared with vanilla pixel or object-level semantics.
To this end, we devise a self-regulated surrogate task to exploit such domain-robust correspondences, by firstly masking nuclei pixels and then resorting to restore the concealed biological contexts.
In the restoration step, the only information available is the characteristics of background tissue structures, as nuclear properties have been covered up. 
On that account, the task becomes predicting the attributes of nuclei according to their surrounding tissue formulation, which provides an elegant solution to learn the relationships between nuclei and tissue.
%Afterwards, the learned relational contexts are distilled to facilitate cross-domain nuclei recognition via the shared backbone.

\begin{comment}
\begin{figure}[!t]
\centerline{\includegraphics[width=0.8\columnwidth]{img-nuclei-mask.png}}
\caption{Examples of the results of image-level nuclei masking and restoration. (a)\,raw images; (b)\,masked images; (c)\,restored images.
It can be noticed that although the restored images inevitably lose some textural details of each individual nucleus, the difference and contrast across nuclei of disparate types are recovered, which is essential for the classification task.}
\label{fig:img-nuclei-mask}
\end{figure}
\end{comment}
\begin{figure*}[!t]
	\centerline{\includegraphics[width=\columnwidth]{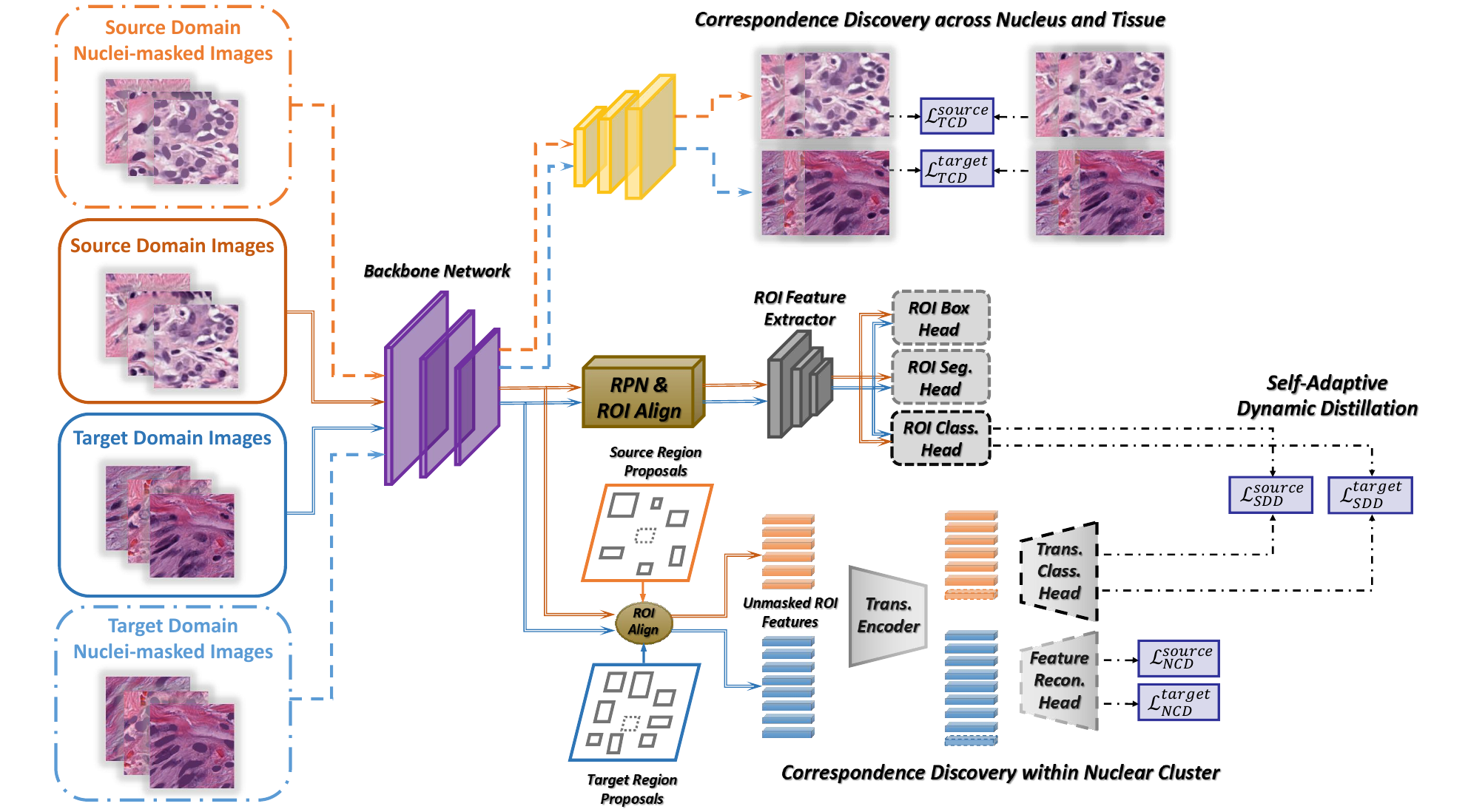}}
	\vspace{-0.5mm}
	\caption{Overview of the proposed approach.
		We aim to learn the implicit correspondences across various biological structures via self-regulated surrogate tasks.
		Specifically, we first perform nuclei masking and then learn to restore the obscured contextual details based on the characteristics of tissue and neighbouring nuclei.
		For correspondence discovery within nuclear cluster, the dotted bounding box and features indicate the location and mask token of the masked nucleus.
		%Trans., Seg., Class., and Recon. in the diagram denote transformer, segmentation, classification, and reconstruction, respectively.
		%M, T, ICD, FCD, and SDD in loss functions denote Mask R-CNN, transformer, image-level correlation discovery, feature-level correlation discovery, and self-adaptive dynamic distillation, respectively.
		%In the figure, only losses related to nuclei classification are marked, and the instance segmentation-related ones are omitted.
		\vspace{-2mm}
	}
	\label{fig:model}
\end{figure*}
Specifically, given an image $\mathbf{I}$ and its nuclei binary mask $\mathbf{\hat{M}}$,
%\;(\emph{i.e.}, a value of 1 indicates nuclei pixels, and a value of 0 indicates background pixels)
we mask the inner details of nuclei regions by replacing all nuclei pixel values with their average value to perform counterfactual intervention: 
$\mathbf{\hat{I}_{mask}} = \mathbf{I} \odot (\mathbbm{1}-\mathbf{\hat{M}}) + \mathtt{avg}(\mathbf{\hat{M}} \odot \mathbf{I}) \odot \mathbf{\hat{M}}$,
where $\odot$ denotes the element-wise product operation.
%For the target domain where ground truth labels are not available, we utilize the pseudo-labels generated with a model trained on the source domain since Fig.\;\ref{fig:tissue-wise} shows that the deep model is robust to domain shifts for class-agnostic segmentation.
%We select the average nuclei pixel values to fill the masked regions to minimize the discrepancy between the raw and masked images.
%Note that nuclei masking works as a pre-processing step and does not require a learning procedure.
%
%\noindent\textbf{Image-level Nuclei Restoration.} \ 
Then, we restore the pixel values of masked nuclei based on the surrounding tissue characteristics and accordingly model the nuclei-tissue relationships.
Considering that the functional types of nuclei are mainly dependent on the locally surrounding tissue structures \cite{ryu2023ocelot}, we construct the pixel-wise restoration decoder with convolutional filter to leverage its strong focus on local contextual information. 
% and circumvent the prohibitive computational burden in adjunct with exhaustive pixel-wise self-attention calculation.
In specific, the masked images $\mathbf{\hat{I}_{mask}}$ are forwarded to backbone $\bm{B}$ for encoding tissue structures, followed by the restoration block $\bm{\tilde{G}}_{\bf{INC}}$ to re-fill those masked regions per pixel.
%A skip connection is further introduced such that the output of the decoder is summed with the masked image to generate 
The final restoration results can be represented with 
$
\mathbf{\hat{I}_{rec}} = \bm{\tilde{G}}_{\bf{INC}}(\bm{B}(\mathbf{\hat{I}_{mask}})) + \mathbf{\hat{I}_{mask}}.
$
%In this way, the purpose of the restoration step is focused on learning critical nuclei-related knowledge, instead of managing to generate realistic background tissue.
%
%\noindent\textbf{Pretext Objective.} \ 
As optimization target of the surrogate restoration task, we impose two training objectives towards disparate yet complementary regularization endpoints:
\begin{equation}
\mathcal{L}_{\rm{TCD}} = \mathbb{E}_{(\mathbf{I}, \mathbf{\hat{I}_{rec}})\,\sim \mathcal{X}_{s/t}}[\mathcal{H}(\mathbf{I}, \mathbf{\hat{I}_{rec}}) +\mathcal{L}_{\rm{perpt}}(\mathbf{I}, \mathbf{\hat{I}_{rec}}; \mathbf{\check{D}})].
\end{equation}
Here $\mathcal{X}_{s/t}$ denote the data distributions of source and target domains.
The first term aims to ensure the pixel-level context coherence after restoration with a matching loss term $\mathcal{H}$.
%To this end, a matching loss term $\mathcal{H}$ is induced to facilitate pixel-level context coherence.
The other perceptual regularization term enforces the generated image to not only locally approach the raw image but also exhibit harmonization and consistency from the global perspective.
%The two constraints with complementary targets cooperatively force high-quality restoration results and encourage the correlations to be well-captured.
An adversarial discriminator\,$\mathbf{\check{D}}$ is trained together with the restoration network for estimating perceptual disparity and deriving $\mathcal{L}_{\rm{perpt}}$ \cite{ganin2015unsupervised}.
%Here $\mathcal{X}_s$ and $\mathcal{X}_t$ are the distributions of the source and target histology image domain, and $\mathcal{L}_{\rm{perpt}}$ indicates the cross entropy loss for adversarial domain discrimination.
%\begin{equation}
%\mathop{\mathtt{max}}\limits_{\mathbf{\check{D}}}\;\mathcal{L}_{\rm{perpt}}(\mathbf{I}, \mathbf{\hat{I}_{rec}}; \mathbf{\check{D}}) = \mathtt{log}\,\mathbf{\check{D}}(\mathbf{I}) + \mathtt{log}\,(\mathbbm{1} - \mathbf{\check{D}}(\mathbf{\hat{I}_{rec}})),
%\end{equation}
% (\mathbf{I}, \mathbf{\hat{I}_{recon}}; \mathbf{\check{D}})

%\begin{small}\begin{equation}\begin{split}
%			\mathop{\mathtt{min}}\limits_{\bm{B}, \bm{\tilde{R}}_{\bf{INR}}}&\;\mathcal{L}_{\rm{perpt}}(\mathbf{I}, \mathbf{\hat{I}_{rec}}; \mathbf{\check{D}}) = \mathcal{L}_{\rm{perpt}}(\mathbf{I}, \mathbf{\hat{I}_{mask}}; \bm{B}, \bm{\tilde{R}}_{\bf{INR}}, \mathbf{\check{D}})\\
%			=& \mathtt{log}\,\mathbf{\check{D}}(\mathbf{I}) + \mathtt{log}\,(\mathbbm{1} - \mathbf{\check{D}}(\bm{\tilde{R}}_{\bf{INR}}(\bm{B}(\mathbf{\hat{I}_{mask}})) + \mathbf{\hat{I}_{mask}})).
%\end{split}\end{equation}\end{small}
%where $\mathcal{L}_{normI}$ and $\mathcal{L}_{adv}$ indicate $L1$ loss and cross entropy loss for adversarial domain discrimination, respectively.

%\subsection{Instance Feature-level Correlation Discovery}
%\label{sec:feature-level}
\noindent\textbf{Correspondence within Nuclear Cluster.} \ 
%\subsubsection{Correspondence within Nuclear Cluster}
Recently, several works \cite{liu2021relational, abousamra2021multi} have underlined the value of considering the community nature of cells,
whereas disregarding the particular value of this property under cross-domain scenarios as the correspondences within nuclei clusters are rooted in the pathogenesis of tumor \cite{rendeiro2021spatial} and could deliver implicit modeling of the invariant factor.
%However, their studies are limited to the uniform distribution scenarios,
%whereas under cross-domain settings, the property is particularly of value as the high-level relationships inside nuclei clusters are rooted in the pathogenesis of tumor \cite{rendeiro2021spatial} and demonstrate stronger robustness to domain shifts.
For example, in all types of organs, epithelial cells tend to cluster together in a ring-like shape \cite{gamper2019pannuke}.
We thereupon devise an instance-level nuclei restoration task to explore the inter-nuclei correspondences.
%Different from image-level nuclei restoration which predicts nuclei pixel values with the characteristics of background tissue, in this module, instance-wise feature masking and restoration are performed based on the nuclei spatial information\;(\emph{i.e.}, object adjacency relationships).
The aim is to predict the attributes of a masked nucleus according to its neighbouring nuclei, formulating a neat recipe to learn the implicit object adjacency relationships.
% and therefore behaves as a pre-text task for inter-nuclei correlation discovery.

%\noindent\textbf{Feature-level Nuclei Masking.} \ 
As indicated in Fig.\;\ref{fig:model}, in each image tile $\mathbf{I}$, we first perform ROI align and thereafter mask the instance-wise feature of the central nucleus.
%We average the coordinates of all proposal centers and select the nucleus whose center is closest to the mean coordinate to conduct masking.
%This is to ensure neighbouring nuclei in every direction are potentially available and get rid of the training unstableness incurred by random selection.
Here, masking means that we do not forward the feature of the masked nucleus to the following network.
We denote the instance-wise feature maps for the masked nucleus and the neighbouring nuclei as $\mathbf{F_{mask}}$ and $\{\mathbf{F_{nbr}}^i\}_{i=1}^{N-1}$, respectively,
where $N$ is the total number of nuclei proposals.
%
%\noindent\textbf{Feature-level Nuclei Restoration.} \ 
The restoration scheme intends to retrieve the masked feature maps via contextual modeling of the neighboring nuclear community.
It is essential to capture the long-range dependencies and spatial relationships among the group of adjacent nuclei so that the attributes of the masked nucleus can be reasonably restored.
In this regard, we develop a ViT \cite{dosovitskiy2021image}-based framework where the high-level correlations across nuclei are exploited with the self-attention mechanism.
%Global pooling is firstly employed for instance-wise feature maps to mitigate the computational burden.
The ROI features of unmasked nuclei are firstly passed through a ViT encoder $\bm{Z}_{\bf{FNR}}$.
A token $\mathbf{TK}$ is then concatenated with the encoded unmasked nuclei to represent the nucleus that is masked and required to be reconstructed.
Positional embeddings $\mathbf{PE}$ are added to incorporate the spatial information of each nucleus into consideration.
Subsequently, those representations are forwarded to the feature reconstruction head $\bm{\tilde{G}}_{\bf{FNC}}$, which is also constructed with a series of transformer blocks, to predict the feature values of the masked nucleus:
\begin{small}\begin{equation}\begin{split}
	\{\mathbf{\hat{F}_{rec}}^i\}_{i=1}^{N} = \bm{\tilde{G}}_{\bf{FNC}}(\mathtt{concat}[\bm{Z}_{\bf{FNR}}(\{\mathbf{F_{nbr}}^i\}_{i=1}^{N-1}), \,\mathbf{TK}] + \mathbf{PE}).
	\end{split}\end{equation}\end{small}%
Then, the restored map for the masked nucleus $\mathbf{\hat{F}_{rec}^{mask}} = \{\mathbf{\hat{F}_{rec}}^i\}_{i=1}^{N}[N]$.
%For the target domain, we utilize the source domain-trained model to generate region proposals and then conduct RoI align and feature masking.
%
%\noindent\textbf{Pretext Objective.} \ 
The training objective of the surrogate task for inter-nuclei correspondence discovery is to restore the original feature maps of the masked nucleus based on neighbouring nuclei characteristics.
Therefore, we apply the matching loss $\mathcal{H}$ to penalize the inconsistency between raw feature maps and the restoration results:
\begin{equation}
\mathcal{L}_{\rm{NCD}} = \mathbb{E}_{(\mathbf{F}|\mathbf{I})\,\sim \mathcal{X}_{s/t}}[\mathcal{H}(\mathbf{F_{mask}},\mathbf{\hat{F}_{rec}^{mask}})].
\end{equation}
%where $\mathbf{F}_{mask}$ and $\mathbf{F}_{rec}^{mask}$ denote the raw and restored instance features of the masked nucleus in image $\mathbf{I}$, respectively.

\subsection{Self-adaptive Dynamic Distillation}
%In Mask R-CNN, classification prediction is made on top of the cropped features, which is dominated by the appearance and texture attributes of each individual nucleus.
%However, several recent works \cite{liu2021relational, abousamra2021multi, hassan2022nucleus} argued that this approach neglects the community nature of cells.
%It fails to leverage the rich contextual information and high-level spatial relationships of surrounding nuclei and processes each nucleus separately.
%There exists a need to enhance Mask R-CNN by going beyond the object itself and harnessing inter-nuclei correlation to perform nuclei type recognition.
%Existing works either rely on fixed data structures (e.g., $3 \times 3$ grids  \cite{liu2021relational}) or pre-defined functions (e.g., K-function \cite{abousamra2021multi}), which is inflexible to model the rich contextual information and high-level spatial relationships of surrounding nuclei.
In standard object recognition pipeline, identification of instance type is dominated by the appearance and texture attributes of each individual object, which are vulnerable to the data distribution biases incurred by variations in imaging protocol and staining procedure \cite{he2017mask, zhao2019object}.
To further leverage the domain-invariant contextual information and spatial relationships of nuclear clusters, we propose to take advantage of transformer's capability to capture high-level implicit correspondences between the group of input nuclei feature representations \cite{han2022survey} and provide adaptive guidance.

Specifically, we reuse the ViT previously deployed to characterize and embed the inter-nuclei correlations with an appended classification head. 
%It is achieved by modeling the set of nuclei cooperatively and measuring their correlations with self-attention.
%We construct a classification head on top of the encoder to identify the category of each nucleus conditioned on the modeled correlations.
%A fully-supervised classification cross-entropy loss is employed for the source domain only, denoted as $\mathcal{L}_{T-class}^{source}$.
%The classification cross-entropy loss for predictions made by Mask R-CNN for the source domain is denoted as $\mathcal{L}_{M-class}^{source}$.
%
Here, the transformer-based branch operates in parallel with the basic convolution-based one.
Given that the two architectures focus on different hierarchies of contextual information\;(\emph{i.e.,} low-level instance-wise attributes and high-level inter-object correspondences, respectively),
performing mutual distillation could strengthen the overall results to transcend object semantics-dominated prediction.
% and approach inter-object correlation-conditioned prediction
%
%A trivial solution is to consider the transformer-based prediction as gold standard and straightforwardly minimize the discrepancy between classification scores across predictions from different branches.
%However, it is noteworthy that the correlation-conditioned prediction is not always optimal.
%We find in experiments\;(see Fig.\;\ref{fig:tissue-wise}) that when performing cross-domain inferences with Mask R-CNN, the performance degradation for connective and inflammatory cells, which are sparsely distributed and possess distinct morphological attributes \cite{gamper2019pannuke}, is relatively moderate.
%It indicates that the object semantics-conditioned prediction could be more reliable for a nucleus that is distant from other nuclei and possesses unambiguous individual characteristics.
Then, we propose to adaptively adjust the trade-offs across two branches for each nucleus according to instance-wise ambiguity.
First, we estimate the uncertainty of model inference with predictive entropy, which measures the quantity of information included in the model’s predictive density function \cite{nair2020exploring}.
%An average prediction $\hat{y}$ is derived based on the $T$ times of stochastic forward passes through the network and averaging the $T$ classification softmax outputs, to present a bayesian
%approximation of dropout \cite{gal2016dropout}.
With the input pairs of neural network denoted as ($\mathbf{x}^*, \mathbf{y}^*$) and its weights denoted by $\mathbf{W}$, 
the approximate predictive distribution is parameterized as:
\begin{equation}
q(\mathbf{y}^* | \mathbf{x}^*) = \int p(\mathbf{y}^* | \mathbf{x}^*, \mathbf{W})q(\mathbf{W})\mathbf{dW},
\end{equation}
where $p(\mathbf{y}^* | \mathbf{x}^*, \mathbf{W})$ represents the predictive distribution, $q(\mathbf{W})$ denotes the approximate variational distribution of $\mathbf{W}$.
The Monte Carlo estimate $\hat{y}$ can be thereafter derived:
\begin{equation}
\text{{$\hat{y} = \mathbb{E}_{q(\mathbf{y}^* | \mathbf{x}^*)}(\mathbf{y}^*)\approx\frac{1}{T}\sum_{t=1}^{T}\hat{\mathbf{y}^*}(\mathbf{x}^*, \mathbf{W}^t),$}}
\end{equation}
where $\hat{\mathbf{y}^*}$ corresponds to the label predictions, $T$ is the number of stochastic forward passes.
Then, the final predictive entropy is obtained by aggregating entropy over all classes
$\mathbf{U} = -\sum_{c=1}^{C}\mathbb{P}(\hat{y}=c)\mathtt{log}\,\mathbb{P}(\hat{y}=c).$
Here $C$ denotes the number of all classes, $\mathbb{P}(\hat{y}=c)$ denotes the probability of $\hat{y}$ belonging to class $c$.
For nuclei recognition, we find that\;(see Fig.\;\ref{fig:tissue-wise}) the cross-domain degradation for connective and inflammatory cells, which are sparsely distributed and possess distinct morphological attributes \cite{gamper2019pannuke}, is relatively moderate.
It indicates that the proposals conditioned on object semantics could be more reliable for a nucleus that is distant from other nuclei and possesses unambiguous individual characteristics.
We thereby consider $\mathbf{U}^M / \mathbf{U}^T$ as the trade-off factor over the convolution- and transformer-based classification modules, with different model tendency assigned to nuclei under divergent spatial distributions.
%where $\mathbf{U}^M$ and $\mathbf{U}^T$ represent the uncertainty of predictions made by convolution- and transformer-based classification heads, respectively.
%When the estimated uncertainty of predictions from the CNN-based classification head is high, it indicates that we should turn to transformer-based prediction, and vice versa.
%
The trade-offs are thereafter adopted to regulate the overall loss function to deliver self-adaptive dynamic guidance:
%Mutual distillation loss is employed on classification softmax outputs to encourage the consistency of two predictions:
\begin{equation}
\text{\resizebox{2.5in}{0.2in}{$\mathcal{L}_{\rm{SDD}} = \mathbb{E}_{\mathbf{I}\,\sim \mathcal{X}_s/\mathcal{X}_t}[ \frac{1}{N}\sum_{i=1}^{N}\frac{\mathbf{U}_i^M}{\mathbf{U}_i^T}\lVert \mathcal{S}_i^M - \mathcal{S}_i^T \rVert; \mathbf{I}],$}}
\end{equation}
where $N$ is the total number of nuclei in image $\mathbf{I}$, $\mathcal{S}^M$ and $\mathcal{S}^T$ are the classification scores of the two model constitutes.

\begin{table*}[!t]
	\centering
	\fontsize{7.5}{8.5}\selectfont
	\begin{threeparttable}
		\caption{Comparison results of our proposed method against other state-of-the-art methods for nuclei classification under three cross-organ settings. X$\rightarrow$Y denotes that the model is trained on data acquired from X organ and then evaluated on Y organ samples. $F$ scores for each class and the class-averaged overall score are reported. Neo., Epi., Con., and Inf. denote neoplastic, epithelial, connective, and inflammatory cells, respectively. Best and second best results are highlighted in \textbf{bold} and \underline{underlined}, respectively.}
		\label{tab:maintable}
		\setlength{\tabcolsep}{0.6mm}{
			\begin{tabular}{c||ccccc|ccccc|ccccc}
				\toprule[0.5mm]
				\multirow{3}{*}{\footnotesize \textbf{Methods}}
				&\multicolumn{5}{c}{\textbf{Breast$\rightarrow$Testis\;(F score)}}
				&\multicolumn{5}{c}{\textbf{Breast$\rightarrow$Thyroid\;(F score)}}
				&\multicolumn{5}{c}{\textbf{Breast$\rightarrow$Bile-duct\;(F score)}}\cr
				\cmidrule(lr){2-6} \cmidrule(lr){7-11} \cmidrule(lr){12-16} 
				&Neo.&Epi.&Con.&Inf.&\textbf{Avg}.&Neo.&Epi.&Con.&Inf.&\textbf{Avg}.&Neo.&Epi.&Con.&Inf.&\textbf{Avg}.\cr
				\midrule[0.3mm]				
				Source-only
				&0.428&0.070&0.529&\underline{0.607}&0.409
				&0.311&0.036&0.445&0.368&0.290
				&0.553&0.000&0.498&0.522&0.393\cr
				\midrule[0.15mm]				
				DA-RCNN \cite{chen2018domain}
				&0.527&0.357&0.579&0.460&0.481
				&0.224&\underline{0.303}&0.381&0.417&0.331
				&0.548&0.024&0.466&0.568&0.401\cr	
				PSA \cite{zheng2020cross}
				&\underline{0.576}&0.256&0.580&0.254&0.417
				&0.386&0.284&0.392&0.401&\underline{0.366}
				&0.566&0.000&0.454&0.462&0.371\cr			
				MGA \cite{zhou2022multi}
				&0.540&0.302&0.556&0.358&0.439
				&0.306&0.116&0.448&0.413&0.321
				&0.535&0.000&0.473&0.526&0.384\cr
				PT-MAF \cite{he2023multi}
				&0.452&0.000&0.547&0.554&0.388
				&0.354&0.000&0.410&0.351&0.279
				&0.516&0.000&0.460&0.446&0.355\cr
				HT \cite{deng2023harmonious}
				&0.518&\underline{0.373}&0.585&0.579&\underline{0.514}
				&0.391&0.264&0.401&0.330&0.347
				&\underline{0.572}&\underline{0.056}&\underline{0.501}&\underline{0.595}&\underline{0.431}\cr
				BAFA \cite{yang2021minimizing}
				&0.535&0.215&0.510&0.572&0.458
				&0.293&0.228&\underline{0.477}&\underline{0.434}&0.358
				&0.558&0.049&0.465&0.590&0.416\cr	
				CAPL-Net \cite{li2022domain}
				&0.551&0.252&\textbf{0.602}&0.364&0.442
				&\underline{0.401}&0.172&0.403&0.389&0.341
				&0.522&0.000&0.481&0.576&0.394\cr
				\midrule[0.15mm]				
				\rowcolor{light-gray} Ours
				&\textbf{0.596}&\textbf{0.521}&\underline{0.594}&\textbf{0.645}&\textbf{0.589}
				&\textbf{0.460}&\textbf{0.359}&\textbf{0.482}&\textbf{0.452}&\textbf{0.438}
				&\textbf{0.627}&\textbf{0.203}&\textbf{0.509}&\textbf{0.615}&\textbf{0.488}\cr	
				\bottomrule[0.5mm]
		\end{tabular}}
	\end{threeparttable}
\end{table*}

\noindent\textbf{Training Pipeline.} \ 
The overall optimization objective is to minimize the aggregated loss:
$\mathcal{L}_{\rm{total}} = \mathcal{L}_{\rm{rec}} + \lambda^*(\mathcal{L}_{\rm{TCD}}+\mathcal{L}_{\rm{NCD}}+\mathcal{L}_{\rm{SDD}}),$
where $\mathcal{L}_{\rm{rec}}$ is the base loss of the recognition model on the source domain, and $\lambda^*$ controls the scaling factors of surrogate tasks.
The supplementary loss terms are aggregated from both the source and target domains.

\section{Experiments and Results}
\subsection{Experimental Setup}
\noindent\textbf{Datasets.} \ 
To verify the effectiveness and general applicability of the proposed method, we perform extensive experiments under various domain shifts incurred by the discrepancy across organs and stains.
For cross-organ adaptation, we leverage four datasets sampled from different organs, \emph{i.e.,} breast, testis, thyroid, and bile-duct, which are retrieved from The Cancer Genome Atlas\;(TCGA) and \cite{gamper2019pannuke}. 
%All images are cropped into $256\times256$ tiles to alleviate the computational cost.
Nuclei are categorized as of neoplastic, epithelial, connective, and inflammatory cells.
%Data augmentation techniques, including scaling, rotation, and flipping, are applied to each dataset.
%The dead cells which only exist in limited types of tissues are not emphasized in the experiments.
The statistics of used data are presented in Table\;\ref{tab:data_statistics}.
%Following the evaluation protocol proposed in \cite{gamper2019pannuke}, we split each dataset into three folds to conduct 3-fold cross-validation.
%In each split, two folds of the dataset are selected to formulate the training set, and the remaining one is left for testing.
%
Then, we further evaluate our method on scenarios where both cross-organ and stain shifts stand out.
Following previous work \cite{yang2021minimizing}, we adopt CoNSep \cite{graham2019hover} and PanNuke \cite{gamper2019pannuke} as the source and target domains.
CoNSep contains histology tiles from one single organ type\;(colon), whilst PanNuke is collected from 19 different organs.
The staining procedures for those datasets are also inconsistent due to different clinical purposes and regulatory requirements across cohorts and countries.
The neoplastic and epithelial classes in PanNuke are merged into one class for label space coherence.

\begin{table}[!t]
	\centering
	\fontsize{8}{9}\selectfont
	\begin{threeparttable}
		\caption{
			Class-wise statistics of nuclei in the used histology image data across different organs.}
		\label{tab:data_statistics}
		\setlength{\tabcolsep}{0.6mm}{
			\begin{tabular}{c|cccc|c}
				\toprule[0.4mm]
				\multirow{3}{*}{\textbf{Organ Type}}&\multicolumn{5}{c}{\textbf{Number of Annotated Nuclei}}\cr
				\cmidrule(lr){2-6} 
				&Neoplastic&Epithelial&Connective&Inflammatory&Total\cr
				\midrule[0.2mm]
				Breast&162,780&109,758&91,053&51,597&415,188\cr
				\midrule[0.05mm]
				Testis&12,021&6,252&10,845&11,160&40,278\cr
				\midrule[0.05mm]
				Thyroid&10,152&13,692&12,828&5,280&41,952\cr
				\midrule[0.05mm]
				Bile-duct&26,460&2,352&23,316&13,995&66,123\cr
				\bottomrule[0.4mm]
		\end{tabular}}
	\end{threeparttable}
	\vspace{-2.4mm}
\end{table}

\begin{table*}[!t]
	\centering
	\fontsize{7.5}{8.5}\selectfont
	\begin{threeparttable}
		\caption{% of our proposed method against other state-of-the-art methods
			Comparison results on cross-organ nuclei instance segmentation. $PQ$ metrics over each class and averaged score are reported. }
		\label{tab:segmentation}
		\setlength{\tabcolsep}{0.6mm}{
			\begin{tabular}{c||ccccc|ccccc|ccccc}
				\toprule[0.5mm]
				\multirow{3}{*}{\footnotesize \textbf{Methods}}
				&\multicolumn{5}{c}{\textbf{Breast$\rightarrow$Testis\;(PQ score)}}
				&\multicolumn{5}{c}{\textbf{Breast$\rightarrow$Thyroid\;(PQ score)}}
				&\multicolumn{5}{c}{\textbf{Breast$\rightarrow$Bile-duct\;(PQ score)}}\cr
				\cmidrule(lr){2-6} \cmidrule(lr){7-11} \cmidrule(lr){12-16} 
				&Neo.&Epi.&Con.&Inf.&\textbf{Avg}.&Neo.&Epi.&Con.&Inf.&\textbf{Avg}.&Neo.&Epi.&Con.&Inf.&\textbf{Avg}.\cr
				\midrule[0.3mm]				
				Source-only
				&0.207&0.029&\textbf{0.344}&0.309&0.222
				&\underline{0.144}&0.032&\textbf{0.320}&0.196&0.173
				&0.295&0.000&\underline{0.288}&0.299&0.220\cr
				\midrule[0.15mm]				
				DA-RCNN \cite{chen2018domain}
				&0.235&0.179&0.323&\underline{0.340}&\underline{0.269}
				&0.095&\underline{0.138}&0.237&\underline{0.271}&0.185
				&0.282&0.011&0.273&0.270&0.209\cr
				MGA \cite{zhou2022multi}
				&0.243&0.150&0.289&0.266&0.237
				&0.121&0.045&0.282&0.257&0.176
				&0.294&0.000&0.263&0.295&0.212\cr
				HT \cite{deng2023harmonious}
				&0.214&\underline{0.207}&0.306&0.313&0.260
				&0.137&0.084&0.262&0.177&0.165
				&\underline{0.309}&\underline{0.032}&\textbf{0.295}&0.283&\underline{0.230}\cr
				BAFA \cite{yang2021minimizing}
				&\underline{0.236}&0.145&0.311&0.332&0.256
				&0.127&0.103&\underline{0.302}&0.264&\underline{0.199}
				&0.297&0.015&0.257&\underline{0.301}&0.218\cr	
				%CAPL-Net \cite{li2022domain}
				%&\textbf{0.255}&0.177&\textbf{0.377}&0.298&0.277
				%&\textbf{0.162}&0.121&0.266&0.217&0.191
				%&0.290&0.000&\textbf{0.329}&\textbf{0.330}&0.237\cr
				\midrule[0.15mm]				
				\rowcolor{light-gray} Ours
				&\textbf{0.252}&\textbf{0.390}&\underline{0.324}&\textbf{0.385}&\textbf{0.338}
				&\textbf{0.159}&\textbf{0.141}&0.297&\textbf{0.284}&\textbf{0.220}
				&\textbf{0.330}&\textbf{0.102}&0.286&\textbf{0.304}&\textbf{0.256}\cr
				\bottomrule[0.5mm]
		\end{tabular}}
	\end{threeparttable}
\end{table*}

\begin{table*}[!t]
	\centering
	\fontsize{8}{9}\selectfont
	\begin{threeparttable}
		\caption{Comparison results for nuclei recognition under both organ and stain shifts. Neo-Epi. denotes the united class for neoplastic and epithelial cells.}
		\label{tab:consep2pannuke}
		\setlength{\tabcolsep}{1.6mm}{
			\begin{tabular}{c||cccc|cccc}
				\toprule[0.5mm]
				\multirow{3}{*}{\footnotesize \textbf{Methods}}
				&\multicolumn{4}{c}{\textbf{CoNSep$\rightarrow$PanNuke\;(F score)}}
				&\multicolumn{4}{c}{\textbf{CoNSep$\rightarrow$PanNuke\;(PQ score)}}\cr
				\cmidrule(lr){2-5} \cmidrule(lr){6-9} 
				&Neo-Epi.&Con.&Inf.&\textbf{Avg}.&Neo-Epi.&Con.&Inf.&\textbf{Avg}.\cr
				\midrule[0.3mm]				
				Source-only
				&0.285&0.278&0.356&0.306
				&0.216&\underline{0.130}&0.149&0.165\cr
				DA-RCNN \cite{chen2018domain}
				&0.316&0.254&\textbf{0.411}&0.327
				&0.227&0.122&0.165&0.171\cr
				MGA \cite{zhou2022multi}
				&\underline{0.370}&0.272&0.359&\underline{0.334}
				&\underline{0.240}&0.117&0.158&\underline{0.172}\cr
				HT \cite{deng2023harmonious}
				&0.348&0.251&0.366&0.322
				&0.195&0.102&0.144&0.147\cr
				BAFA \cite{yang2021minimizing}
				&0.261&\underline{0.280}&\underline{0.403}&0.314
				&0.183&0.120&\textbf{0.174}&0.159\cr	
				CAPL-Net \cite{li2022domain}
				&0.335&0.264&0.332&0.310
				&0.231&0.113&0.155&0.167\cr	
				\rowcolor{light-gray} Ours
				&\textbf{0.404}&\textbf{0.300}&0.401&\textbf{0.368}
				&\textbf{0.258}&\textbf{0.149}&\underline{0.166}&\textbf{0.191}\cr
				%\bottomrule[0.5mm]
		\end{tabular}}
	\end{threeparttable}
	\begin{threeparttable}
		\setlength{\tabcolsep}{1.65mm}{
			\begin{tabular}{c||cccc|cccc}
				\toprule[0.5mm]
				\multirow{3}{*}{\footnotesize \textbf{Methods}}
				&\multicolumn{4}{c}{\textbf{PanNuke$\rightarrow$CoNSep\;(F score)}}
				&\multicolumn{4}{c}{\textbf{PanNuke$\rightarrow$CoNSep\;(PQ score)}}\cr
				\cmidrule(lr){2-5} \cmidrule(lr){6-9} 
				&Neo-Epi.&Con.&Inf.&\textbf{Avg}.&Neo-Epi.&Con.&Inf.&\textbf{Avg}.\cr
				\midrule[0.3mm]				
				Source-only
				&0.796&0.639&0.590&0.675
				&\underline{0.305}&\underline{0.213}&0.312&0.276\cr
				DA-RCNN \cite{chen2018domain}
				&0.819&0.588&0.603&0.669
				&0.289&0.174&0.335&0.265\cr
				MGA \cite{zhou2022multi}
				&0.774&0.609&0.602&0.661
				&0.266&0.185&0.302&0.251\cr
				HT \cite{deng2023harmonious}
				&\underline{0.830}&0.647&0.611&\underline{0.696}
				&0.301&0.195&0.318&0.271\cr
				BAFA \cite{yang2021minimizing}
				&0.793&\underline{0.665}&\textbf{0.626}&0.695
				&0.298&0.207&\underline{0.349}&\underline{0.285}\cr	
				CAPL-Net \cite{li2022domain}
				&0.768&0.574&0.606&0.649
				&0.273&0.178&0.320&0.257\cr	
				\rowcolor{light-gray} Ours
				&\textbf{0.856}&\textbf{0.696}&\underline{0.618}&\textbf{0.723}
				&\textbf{0.356}&\textbf{0.252}&\textbf{0.352}&\textbf{0.320}\cr
				\bottomrule[0.5mm]
		\end{tabular}}
	\end{threeparttable}
\end{table*}

\noindent\textbf{Implementation Details and Evaluation Metrics.} \ 
Following previous works \cite{hsu2021darcnn}, we adopt Mask R-CNN \cite{he2017mask} as the base model.
The matching loss $\mathcal{H}$ between a pair of inputs is implemented with $L_1$ regularization term.
%Ablation over different total loss weighting terms $\lambda^*$ is presented in the supplementary.
For nuclei masking on the target domain, we utilize the box proposals and masks generated with a model trained on the source domain since Fig.\;\ref{fig:tissue-wise} shows that deep model is robust to domain shifts for class-agnostic segmentation.

For evaluation on the classification task, 
we follow previous works \cite{graham2019hover} and adopt the $F$ score to measure the performance of nuclei classification: $F = \frac{TP}{TP + FP + FN}.$
%\begin{equation}
%F = \frac{TP}{TP+\alpha_1 \cdot FP+\alpha_2 \cdot FN},	
%\end{equation}
%Here $TP$, $FP$, and $FN$ denote the number of true positives, false positives, and false negatives, respectively.
%$\alpha_1$ and $\alpha_2$ are weighting parameters.
More weights are assigned to FP and FN compared with normal $F_1$ score to impose emphasis on false classification results.
For instance segmentation, 
we use panoptic quality\;(PQ) score \cite{gamper2019pannuke} for quantitative evaluation.
Both $F$ and $PQ$ scores are computed for each class and then averaged to demonstrate the overall performance.
%Besides, instead of calculating the scores for each image patch individually and then averaging all the results, we directly accumulate the statistics over all images and then compute the scores for only one time by considering all the testing images as a whole.
%This is to neutralize the impacts of extreme cases where only very few nuclei are contained in a image patch.

\subsection{Comparison with State-of-the-Art Methods}
We compare our proposed method against the state-of-the-art UDA object recognition methods, including DA-RCNN \cite{chen2018domain}, PSA \cite{zheng2020cross}, MGA \cite{zhou2022multi}, PT-MAF \cite{he2023multi}, HT \cite{deng2023harmonious}, BAFA \cite{yang2021minimizing}, and CAPL-Net \cite{li2022domain} to justify its effectiveness.
The results when the source domain-trained model is adopted straight for evaluations on the target domain are also presented for reference.
For fair comparisons, those methods are implemented with the same backbone architecture and training settings\;(batch size, learning rate, \emph{etc.}) as ours.
%including ones designed specifically for nuclei and ones proposed for general objects.
We report the quantitative comparison results for each type of nuclei and the averaged value over all classes under a 3-fold cross-validation setting.
%Those results are calculated with a 3-fold cross-validation setting.
Paired t-tests are also conducted between our method and the others on class-averaged overall scores.
The resulting \textbf{p-values} for all tests are \textbf{below 0.05}, indicating the proposed method significantly outperforms the approaches in comparison.
%Details of the statistical analysis are illustrated in Section \ref{sec:statistics}.

\begin{figure}[!t]
	\centerline{\includegraphics[width=0.95\columnwidth]{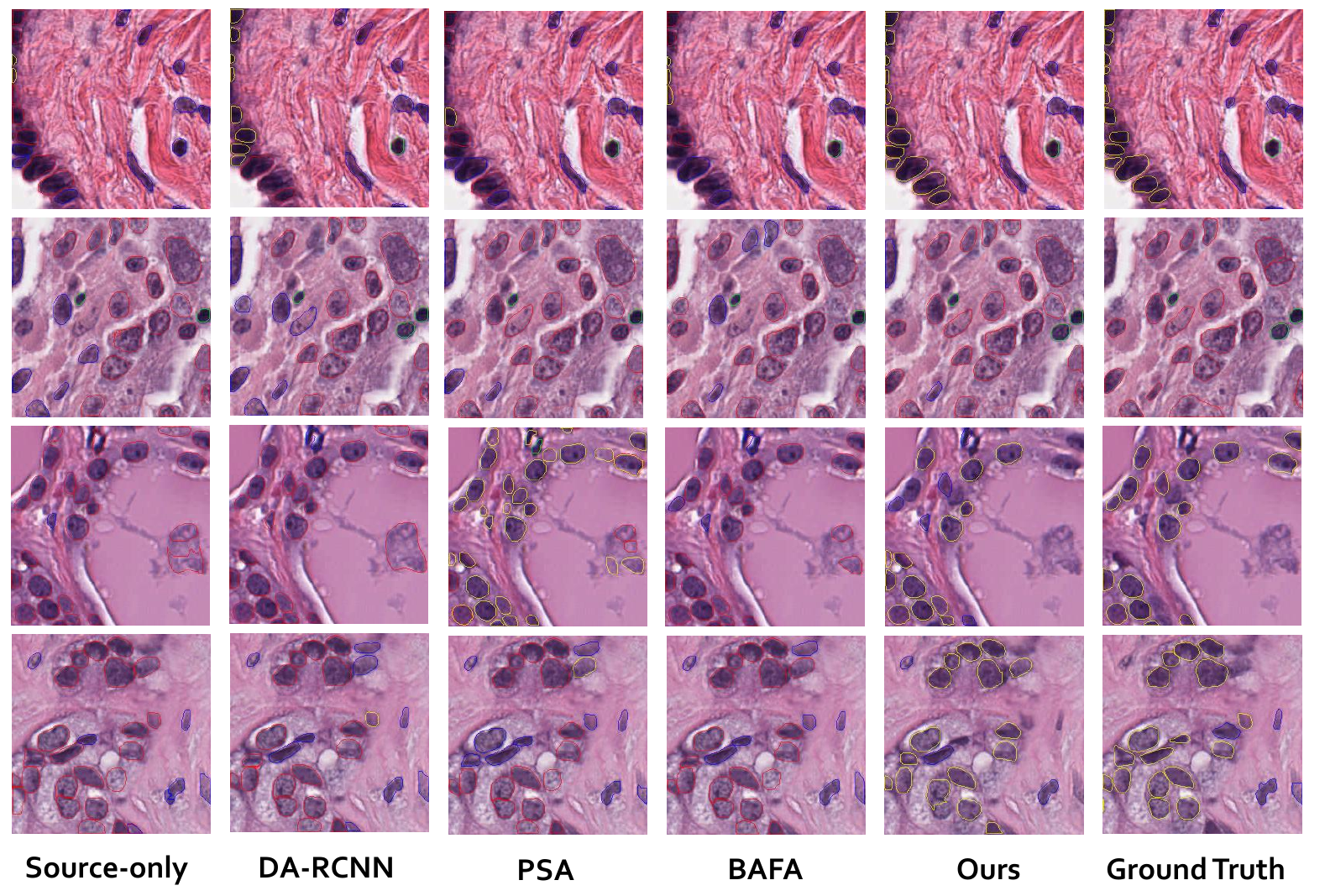}}
	\caption{Qualitative comparison of nuclei recognition results.
		Images in the top two rows are from the testis, whereas images in the third and fourth rows are from the thyroid and bile-duct, respectively.
		In each sub-figure, red, yellow, blue, and green contours correspond to the nuclei of neoplastic, epithelial, connective, and inflammatory cells, respectively.
		%Results for Breast→Testis and Breast→Thyroid settings are presented. 
		%The box plots and p-values confirm that the improvements our method attains are statistically significant.
	}
	\label{fig:mainvisresults}
	\vspace{-2.5mm}
\end{figure}

In Table\;\ref{tab:maintable}, we present the results for classification task under cross-organ domain shifts.
It is observed that our method achieves significant improvements in terms of class-averaged scores in all the three adaptation scenarios.
% and even achieves superior performance comparable to the fully-supervised upper bound.
The advancement can be attributed to our proposed method's capacity of identifying nuclei with ambiguous semantics and bypassing performance degradation incurred by inaccurate category pseudo-labels.
The results are further discussed in Section\;\ref{sec:Discussion}.
%Further discussions about the results are presented in Appendix 3.2.
In Table\;\ref{tab:segmentation}, the quantitative performance of category-wise instance segmentation is presented.
With the attained high-quality nuclei type identification results, our method concurrently yields appealing accuracy regarding class-wise and -averaged PQ.
%It can be observed that the improvements of UDA methods against source-only baseline is relatively modest compared with classification. The reason behind this is that with multiple new loss items introduced in training, the influence of loss items for segmentation is neutralized, which consequently leads to the deterioration of segmentation quality. 
%Nevertheless, since the accuracy of nuclei classification is improved remarkably, our method could obtain high-quality class labels and therefore yield a higher score regarding class-wise and -averaged PQ.
%The qualitative comparison results are appended in the supplementary. 
Additionally, as depicted in Fig.\;\ref{fig:mainvisresults}, most competing methods fail to recognize epithelial cells and undesirably categorize them into neoplastic ones.
On the contrary, our proposed method successfully distinguishes the two types of nuclei with very similar visual attributes.

To further verify the generalizability of our method, we evaluate its efficacy under both organ and stain shifts.
Following \cite{yang2021minimizing}, we adopt CoNSep and PanNuke to construct the adaptation benchamark and perform bi-directional experiments.
We adopt the evaluation metrics in \cite{yang2021minimizing} to jointly consider the performance of nuclei detection and classification tasks.
PQ scores are also presented for segmentation quality evaluation.
As shown in Table\;\ref{tab:consep2pannuke}, we compare our method with the state-of-the-art UDA object recognition approaches.
%It is noticed that the adaptation modules of Hsu \emph{et al.} \cite{hsu2020progressive} are detrimental to the cross-domain evaluation accuracy.
%This is due to the complex data distribution of the target domain and the resulting inferior image-level alignment results.
The empirical improvements of our method in the sophisticated cross-domain setting are consistent with previous experiments and observations, which substantiates the effectiveness and robustness of our method against various types of domain shifts in histology data.

\subsection{Ablation Study}
\label{sec:ablation}
To validate the efficacy of key components in the proposed method, we perform ablation studies on the classification and category-wise instance segmentation tasks by evaluating with several variants of the method.
The corresponding quantitative comparison results are reported in Table\;\ref{tab:ablation}, where MD denotes the mutual distillation across architectures and SA is the estimated trade-offs for self-adaptive dynamic distillation.
It is remarked that all the components have positive impacts on improving the overall classification accuracy.
In specific, solely employing TCD or the combination of NCD and MD can already lead to competitive results.
It justifies the importance to explore the implicit biological correspondences for cross-domain nuclei recognition.
By integrating those constituents together, we reach peak performance.
Moreover, the employment of instance-adaptive guidance further boosts the $F$ score by around $3\%$.
%Those results clearly demonstrate the contributions of the proposed components.
%\begin{comment}
For instance segmentation,
we observe that NCD tends to have a negative impact on the overall accuracy. 
The reason could be that in NCD, we use the global pooling strategy, which inevitably discards the fine-grained spatial information.
In contrast, TCD exhibits beneficial effects on segmentation.
It is in virtue of the design that when performing image-level nuclei masking, we keep all the spatial details and introduce reliable masks, which subsequently serves as a guidance for segmentation.
%\end{comment}
\begin{table}[!t]
	\centering
	\fontsize{8}{9}\selectfont
	\begin{threeparttable}
		\caption{
			Ablation study to verify the efficacy of key components in our method. 
			The class-averaged overall $F$ and $PQ$ scores are presented for each case.
			Tes., Thyr. and Bile. stand as the abbreviations for testis, thyroid, and bile-duct, respectively.
			$\checkmark$ marks indicate the utilized modules.
			The best results are highlighted in \textbf{bold}.}
		\label{tab:ablation}
		\setlength{\tabcolsep}{0.75mm}{
			\begin{tabular}{c||cccc||cc|cc|cc}
				\toprule[0.3mm]
				& \textbf{TCD} & \textbf{NCD} & \textbf{MD} & \textbf{SA} & \textbf{Tes.\,(F)} & \textbf{Tes.\,(PQ)} & \textbf{Thyr.\,(F)} & \textbf{Thyr.\,(PQ)} & \textbf{Bile.\,(F)} & \textbf{Bile.\,(PQ)} \cr
				\midrule[0.15mm]				
				\multirow{7}{*}{\textbf{Settings}}
				&-- &-- &-- &-- &0.409 &0.222 &0.290 &0.173 &0.393 &0.220 \cr
				& $\checkmark$ &-- &-- &-- &0.446 &0.232 &0.326 &0.189 &0.427 &0.227 \cr
				& $\checkmark$ & $\checkmark$&-- &-- &0.472 &0.240 &0.346 &0.180 &0.431 &0.214 \cr
				&-- & $\checkmark$& $\checkmark$ &-- &0.508 &0.256 &0.382 &0.197 &0.465 &0.234 \cr
				& $\checkmark$ & $\checkmark$& $\checkmark$ &-- &0.557 &0.294 &0.401 &0.215 &0.473 &0.250 \cr
				& $\checkmark$ &-- & $\checkmark$ & $\checkmark$ &0.554 &0.327 &0.417 &\textbf{0.231} &0.448 &0.247 \cr
				& $\checkmark$ & $\checkmark$& $\checkmark$ & $\checkmark$ &\textbf{0.589} &\textbf{0.338} &\textbf{0.438} &0.220 &\textbf{0.488} &\textbf{0.256} \cr
				\bottomrule[0.3mm]
		\end{tabular}}
	\end{threeparttable}
\end{table}

\begin{wrapfigure}{r}{0.45\columnwidth}
	\centering
	\vspace{-4mm}
	\includegraphics[width=0.45\columnwidth]{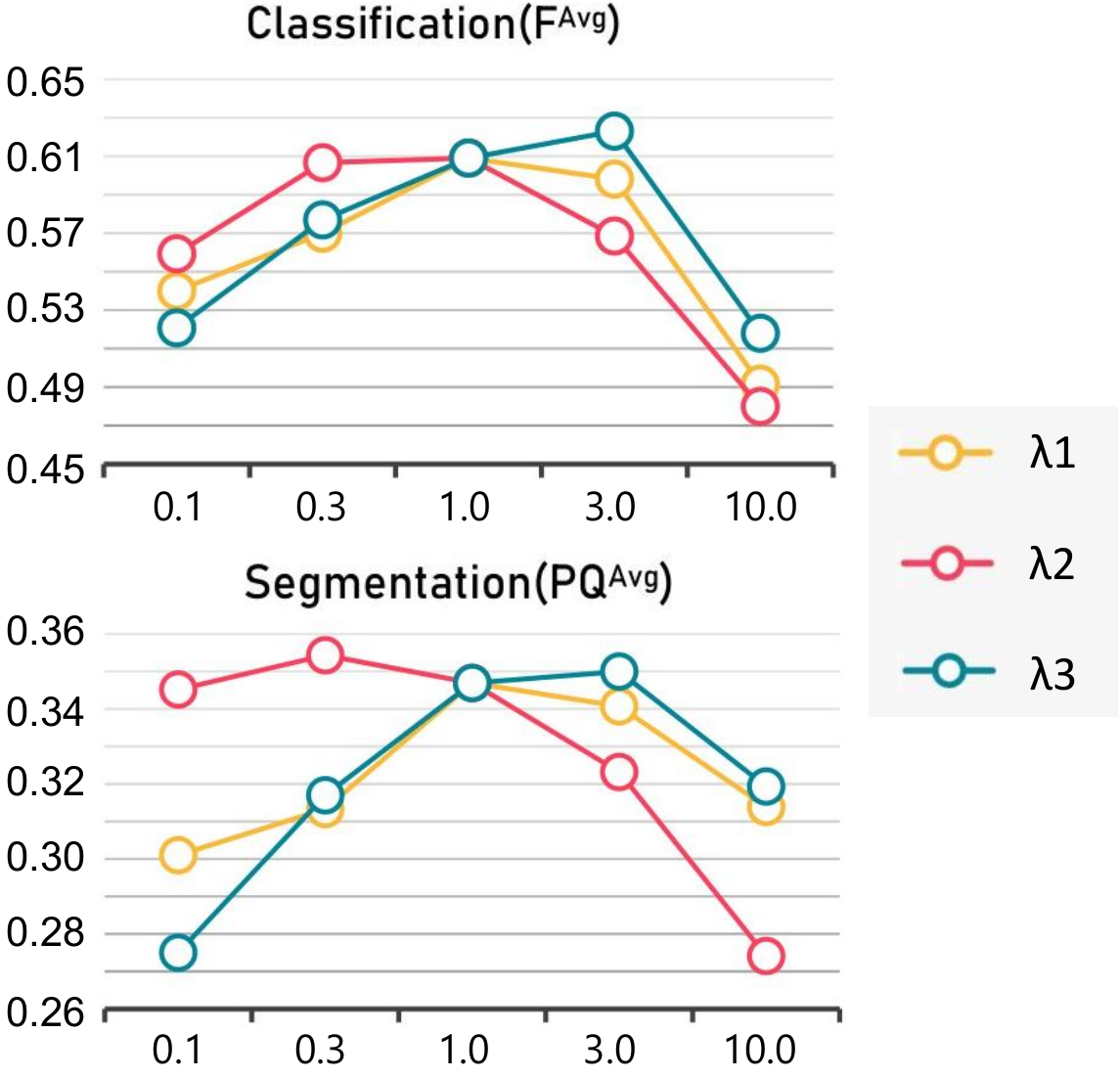}
	\caption{
		Sensitivity analysis of loss weighting items on the Breast$\rightarrow$Testis setting. 	
	}
	\label{fig:sensitivity}
	\vspace{-4mm}
\end{wrapfigure}

To investigate the beneficial impact of our proposed method in greater detail, 
%further ablation study is performed to provide more rigorous quantitative supports.
we perform sensitivity analysis on the loss weighting terms $\lambda^*=\{\lambda_1,\lambda_2,\lambda_3\}$ set, which correspond to the scaling factors of tissue correspondence discovery, nuclear correspondence discovery, and self-adaptive dynamic distillation, respectively.
The results are depicted in Fig.\;\ref{fig:sensitivity}.

For each weighting item, we scale it by a factor of 0.1, 0.3, 1.0, 3.0, and 10.0, respectively, and keep other items fixed.
The corresponding quantitative evaluation results for classification and category-aware instance segmentation are reported.
All the experiments are conducted on the Breast$\rightarrow$Testis setting.
It can be observed that our main setting\;(\emph{i.e.}, when all scale factors equal to 1.0) achieves superior performance compared with most adjusted settings.
In addition, when decreasing the influential factor of $\lambda_3$, there exists a noticeable performance drop for both tasks.
This finding is consistent with the results of ablation studies and substantiates the importance of the proposed instance-adaptive dynamic distillation strategy.
On the other hand, increasing the influential factor of $\lambda_2$ has detrimental effects on instance segmentation, which is mainly attributed to the introduced global pooling strategy.

\subsection{Discussions}
\label{sec:Discussion}
\noindent\textbf{Identification of Nuclei with Ambiguous Semantics.} \ 
Regarding the results in Table\;\ref{tab:maintable}, the advancements of our method can be mainly attributed to its capability for precisely identifying neoplastic and epithelial cells.
\begin{figure}[!t]
	\centerline{\includegraphics[width=0.8\columnwidth]{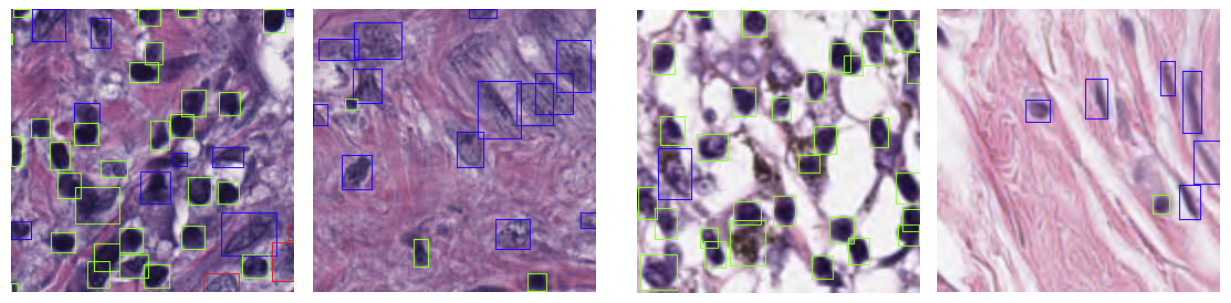}}
	\caption{Examples of H\&E-stained histology image regions. 
		In each sub-figure, blue and green rectangles correspond to the nuclei of connective and inflammatory cells, respectively.}
	\label{fig:con_inf}
	\vspace{-2.5mm}
\end{figure}
Different from connective and inflammatory cells which possess distinct individual shape and texture characteristics\;(\emph{i.e.}, the connective cell typically exhibits a flat polygon pattern in terms of geometric shape and the inflammatory cell is much darker than others in color space, as showcased in Fig.\;\ref{fig:con_inf}), the ambiguity between neoplastic and epithelial cells makes them indistinguishable for object semantics-conditioned model under cross-domain scenarios.
It is challenging to distinguish those two types of cells solely based on their appearance attributes.
To this end, existing methods \cite{chen2018domain, zheng2020cross} which perform UDA with object-wise alignment struggle to find the decision boundary to separate neoplastic and epithelial cells.
In contrast, by exploiting the informative correlations across biological structures, which demonstrate stronger visual contrast, our proposed method significantly lifts the performance to distinguish those two types of cells with analogous morphological traits.
%The $F$ scores regarding epithelial cells are improved by more than $14\%$, $6\%$, and $18\%$ in the three cross-tissue settings, respectively.

%Psuedo label
\noindent\textbf{Sidestepping Reliance on Biased Pseudo-labels.} \ 
With respect to methods built upon category pseudo-labels \cite{zheng2020cross, yang2021minimizing, zhou2022multi}, they bring relatively limited empirical gains for cross-domain nuclei recognition.
For example, in the Breast$\rightarrow$Testis setting, the overall average $F$ scores of those methods are exceeded by the ones that do not depend on category pseudo-labels by almost $5\%$.
%And in particular for the Breast$\rightarrow$Bile-duct setting, \cite{zheng2020cross} only reaches a classification accuracy even $3\%$ lower than the source-only baseline.
The degradation is on the pitfalls of the successive error accumulation caused by biased category pseudo-labels, which is inevitable considering the drastic model collapse across sampling organs and staining protocols.
%Given the substantial discrepancy between the same type of nuclei from different domains, it is infeasible to acquire pseudo-labels of high quality.
In this regard, with specifically devised surrogate tasks as bridges for model transfer across domains, our proposed method gets rid of the self-training scheme and the reliance on category pseudo-labels, which contributes to remarkable improvements in consequence.
%Although segmentation pseudo-labels are utilized in our method, they are relatively reliable since the performance drop for binary DICE and PQ in Table\;\ref{tab:degradation} is moderate.

\begin{figure}[!t]
	\centerline{\includegraphics[width=0.9\columnwidth]{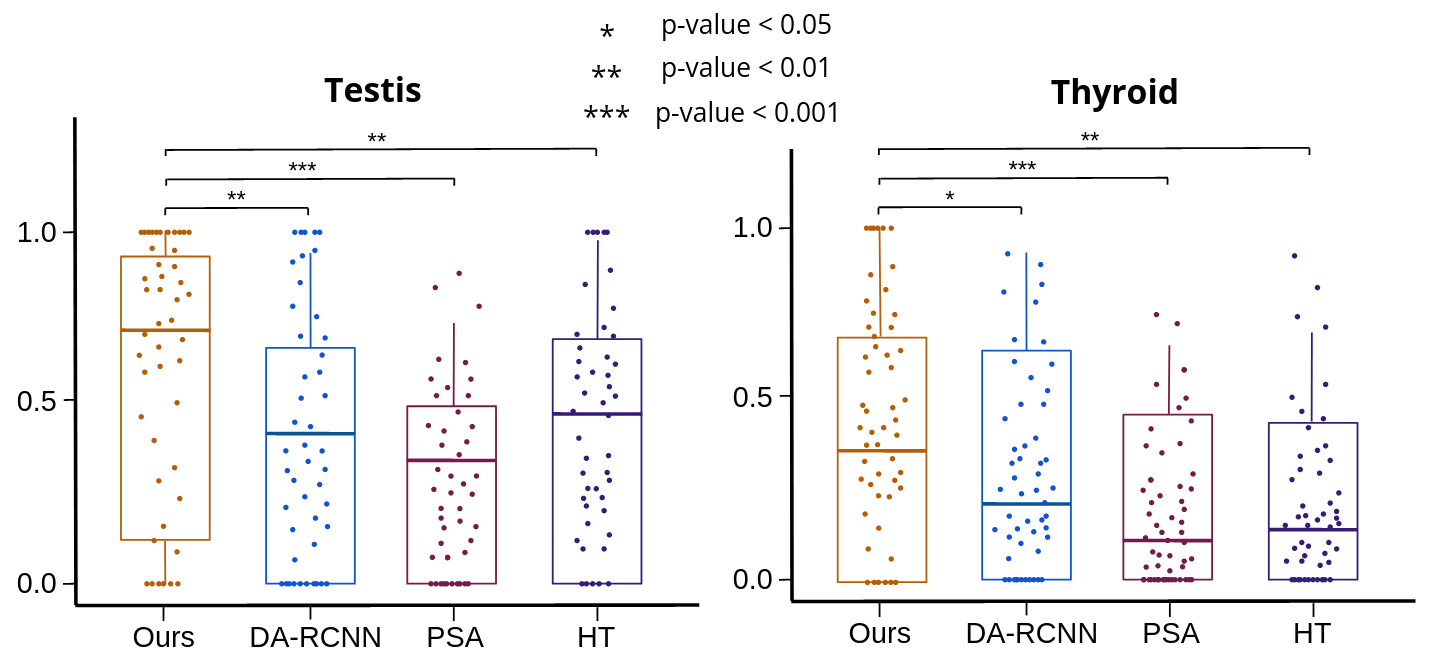}}
	\vspace{-2mm}
	\caption{Statistical analysis on nuclear type recognition results for morphologically ambiguous epithelial cells with paired t-tests. 
		%Results for Breast→Testis and Breast→Thyroid settings are presented. 
		%The box plots and p-values confirm that the improvements our method attains are statistically significant.
	}
	\label{fig:statistics}
	%\vspace{-4mm}
\end{figure}

\noindent\textbf{Statistical Analysis.} \ 
The key advancement of our method, as previously discussed, is its capability to distinguish nuclei with ambiguous morphological characteristics.
%\;(\emph{e.g.}, neoplastic and epithelial cells). 
We therefore conduct detailed statistical analysis on the nuclear classification results regarding epithelial cells. 
%To this end, we calculate the F score for each image individually if there exists at least one nucleus of epithelial cells. 
The resulting box plots and p-values shown in Fig.\;\ref{fig:statistics} substantiate the statistical significance of our achieved improvements.

\section{Conclusion}
%In this work, we propose a novel framework to explore the domain-robust nuclei-background tissue and nuclei-nuclei relationships for cross-tissue nuclei type recognition.
In this work, we propose a holistic framework to facilitate cross-domain cellular nuclei recognition via exploitation of implicit biological relationships at image and instance feature levels.
Additionally, we devise self-adaptive dynamic distillation to further leverage the rich relational contexts inherently present in nuclear communities with instance-aware trade-offs across model architectures.
Experiments on several cross-domain settings with organ and stain shifts demonstrate that our method addresses the common issues existing in the state-of-the-art UDA object recognition approaches and achieves compelling performance.
%In addition, through ablation study, the effectiveness of each component is confirmed.
%This work still has limitations that it obeys the assumption of UDA and requires similar label definitions between the source and target domain.
In future work, we will investigate more challenging yet practical domain adaption scenarios when the cross-domain shifts concurrently stand for class distributions.
%, to enable taxonomy adaptive model generalization.

%\section*{Acknowledgements}
%Please insert your acknowledgments here.

\bibliographystyle{splncs04}
\bibliography{egbib}
\end{document}